\documentclass[12pt]{article}
\usepackage{pdproc} 
\usepackage{epsfig}
\usepackage{latexsym}

\begin{document}

\renewcommand{\thefootnote}{\alph{footnote}}

\newcommand{\lwig}{\mbox{\,\raisebox{.3ex}
    {$<$}$\!\!\!\!\!$\raisebox{-.9ex}{$\sim$}\,}}
\newcommand{\gwig}{\mbox{\,\raisebox{.3ex}
    {$>$}$\!\!\!\!\!$\raisebox{-.9ex}{$\sim$}}\,}
\newcommand{\lambdabar}{{\hbox{$\lambda$\kern-1.ex\raise+0.45ex\hbox{--}}}}

\title{
{\normalsize\rm \rightline{DESY 05-045}}
\vskip 1cm 
 HOW TO DETECT BIG BANG RELIC NEUTRINOS?\footnote{Talk at  
 XI International Workshop on ``Neutrino Telescopes'', Feb 22-25, 2005, Venice, Italy.}
       }

\author{Andreas Ringwald}

\address{Deutsches Elektronen-Synchrotron DESY\\
 Notkestra\ss e 85, D--22607 Hamburg, Germany\\
 {\rm E-mail: Andreas.Ringwald@desy.de}}

\abstract{
The existence of big bang relic neutrinos---exact analogues of the big bang relic 
photons comprising the cosmic microwave background radiation---is a basic prediction 
of standard cosmology. At present, the observational evidence 
for their existence rests entirely on cosmological measurements, such as the light elemental abundances, 
anisotropies in the cosmic microwave background, and the large-scale matter power spectrum. 
In this review, we concentrate on the prospects of more direct, weak interaction based relic neutrino 
detection techniques which are sensitive to the cosmic neutrino background near the 
present epoch and in our local neighborhood in the universe. In this connection, we emphasize
the necessity to take into account the gravitational clustering of the relic neutrinos in 
the cold dark matter halos of our Milky Way and nearby galaxy clusters.     
	 }
   
\normalsize\baselineskip=15pt

\section{\label{intro}Introduction}

Over the last few years, significant advances were made in observational cosmolo\-gy\cite{Eidelman:2004wy}. 
The position of the first Doppler peak in recent 
measurements of the cosmic microwave background (CMB) radiation  
strongly suggests that the universe is spatially flat. 
Observations of Type Ia supernovae (SN) and the large scale structure (LSS) 
favor a universe with present energy fractions of 70 \% of dark energy---accounting for the observed 
accelerating expansion of the universe---and about 25 \% of cold dark matter---the corresponding particles 
being non-relativistic already at the time of recombination.     
Successful big-bang nucleosynthesis (BBN) of the
light elements requires that about 5 \% of the energy content of the universe is in the form of 
ordinary baryonic matter. All in all, we have now a pretty good knowledge of the
cosmic recipe (cf.~Table~\ref{tab:recipe}). A lot of further observational 
and theoretical effort will go into this field to further substantiate 
and explain these cosmological findings, and, in particular, to clarify the 
nature of dark energy and dark matter.  

\begin{table}
\caption{The cosmic recipe.}\label{tab:recipe}
  \small 
\begin{center}
\begin{tabular}{|l|c|c|c|r|l|}
\hline
Material & Particles & $\langle E\rangle$ or $m$ & $ N $ & { $\langle\rho\rangle/\rho_c$} & Obs. Evid. \\
\hline
\hline
{ Radiation} &  $\gamma$ & $0.1$~meV & $10^{87}$ & { 0.01\,\%} & CMB \\
          &                         &              &           &           & \\
\hline
{ Hot Dark} & & $> 0.04$~eV  &  & { $> 0.1$\,\%}   & BBN  \\
{ Matter}             & { Neutrinos} &  & $10^{87}$ &  &  CMB \\
  &       &  $< 0.6$~eV              &           & { $< 2$\,\%}        & LSS \\
\hline
{ Ordinary} & &  &  &   & BBN  \\
{ Matter} &   $p,n,e$      &  MeV-GeV       &  $10^{78}$         &  { 5\,\%}     & CMB      \\
   &       &         &           &       &   \\
\hline 
{ Cold Dark} & { WIMPs?} & $\gwig 100$~GeV & $\lwig 10^{77}$ &  & LSS \\
{ Matter}    &                         &  &   & { 25\,\%}      & CMB  \\
                   &    { Axions?}    &   $\lwig $~meV    &    $\gwig 10^{91}$          &       & \\
\hline
{ Dark} &  &  & &  & SN  \\
{ Energy} & ?  & $10^{-33}$~eV              & ?   &    { 70\,\%}          & CMB  \\
                &   &               &   &              &  \\
\hline
\end{tabular}
\end{center}
\end{table}

Along with the CMB, standard big bang theory predicts the existence of a 
cosmic neutrino background (C$\nu$B). Presently, the evidence for the existence 
of the relic neutrinos rests on the aforementioned cosmological measurements, i.e. the 
light elemental abundances, CMB anisotropies, and the large-scale matter power 
spectrum\cite{Hannestad:2004nb,Pastor:2005proc}. Note, however, that all these measurements probe 
only the presence of the relic neutrinos at early stages in the cosmological 
evolution, and this often in a rather indirect way. It is therefore natural to ask: 
what are the prospects of a more direct, weak interaction based relic neutrino 
detection, sensitive in particular to the C$\nu$B in the present epoch\cite{Gelmini:2004hg}?      
After all, among the known elementary particles, neutrinos are one of the most abundant particles
in the present universe, falling second only to the relic photons (cf.~Table~\ref{tab:recipe}). 

\section{\label{howmany}How Many? How Fast?}

In order to design a direct, weak interaction based detection experiment, 
a precise knowledge of the phase space distribution of the relic neutrinos is indispensable.
In this section, we will review recent determinations of this distribution\cite{Singh:2002de,Ringwald:2004np}. 

\begin{figure}
\begin{center}
         \mbox{\epsfig{file=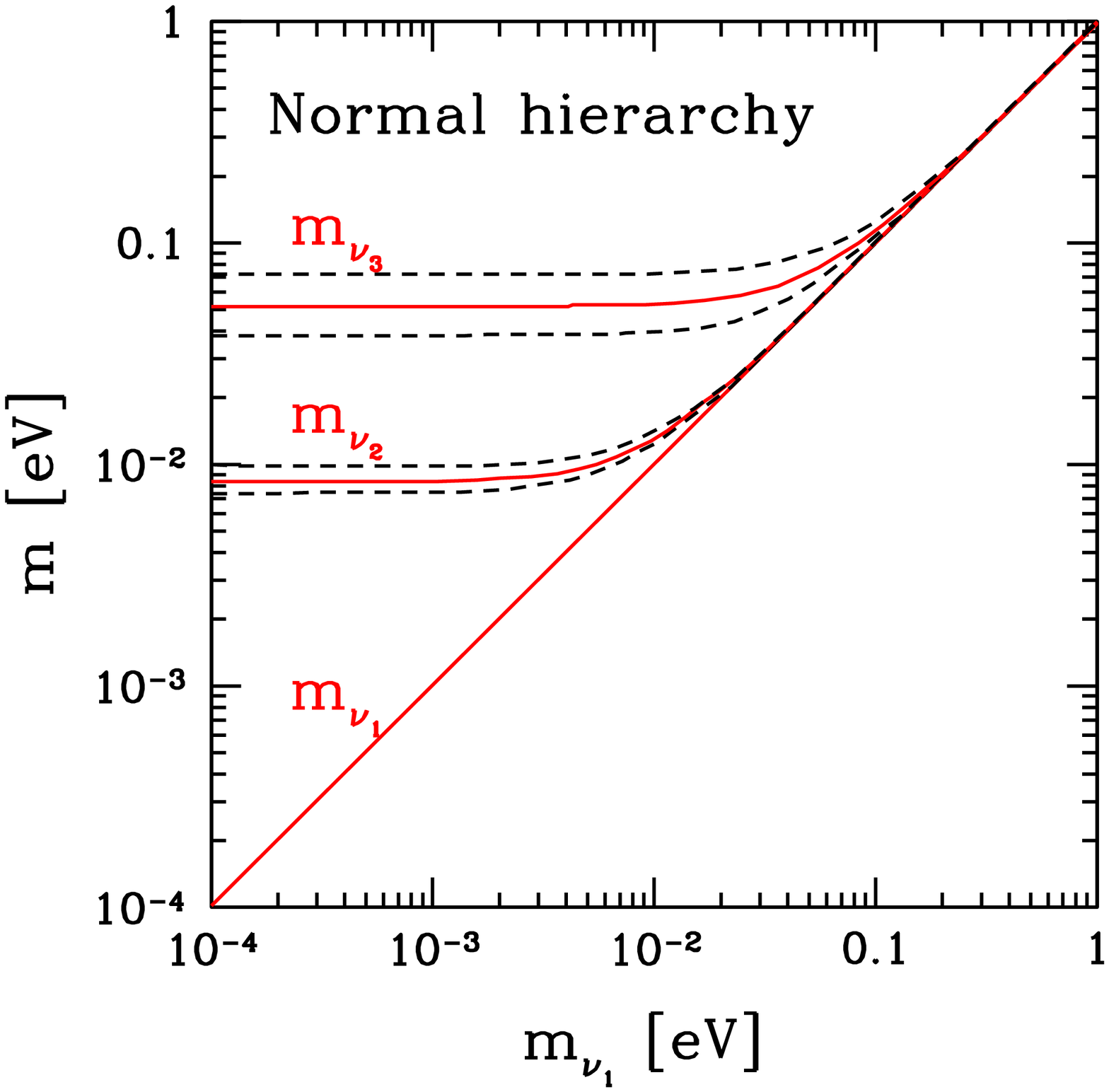,width=7.5cm,clip=}}
\hfill
         \mbox{\epsfig{file=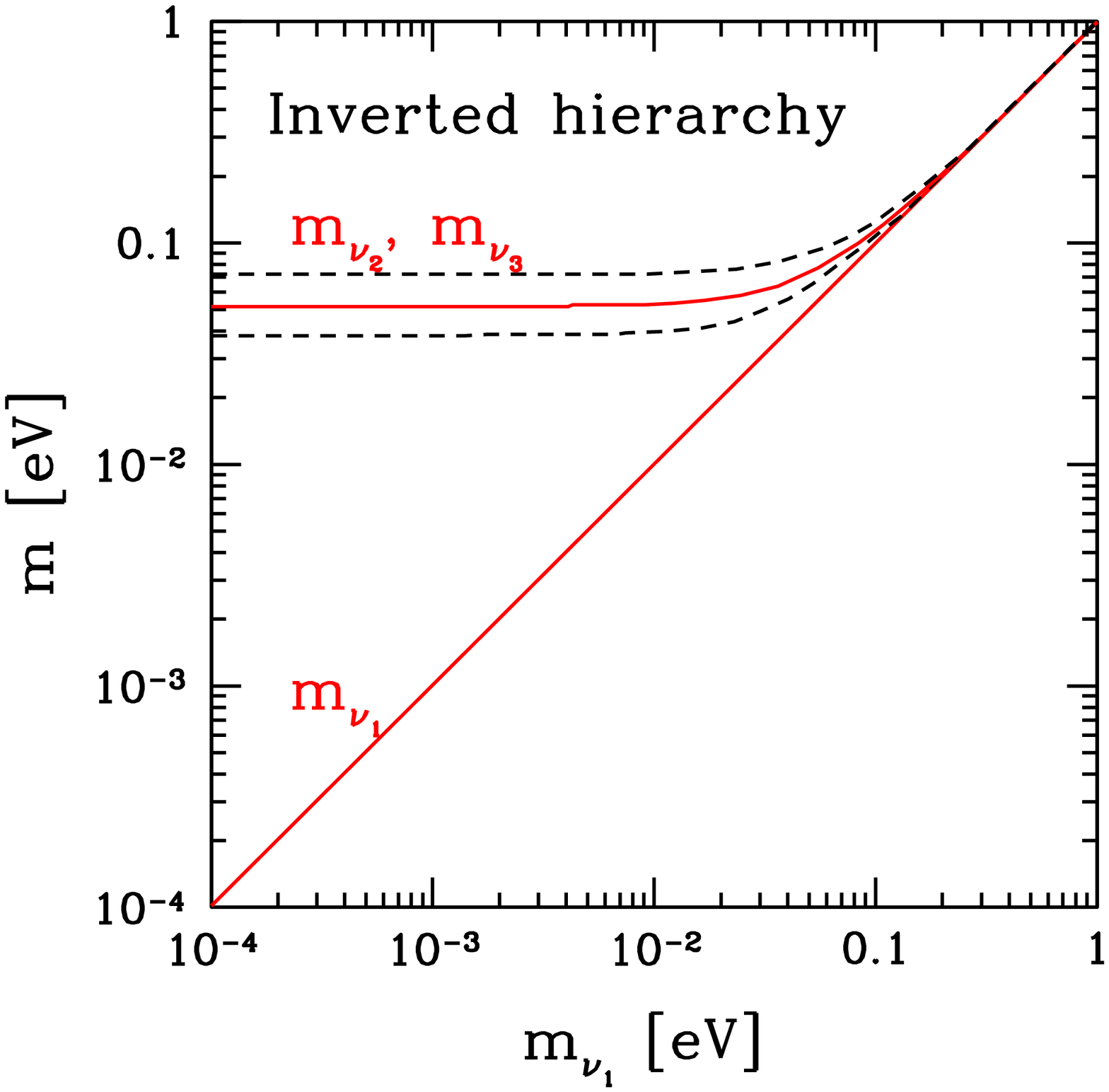,width=7.5cm,clip=}}
\end{center}
\caption[]{Present knowledge about the neutrino mass spectrum\cite{Eberle:2004ua}, 
as a function of the mass of the lightest neutrino, $m_{\nu_1}$.
In the normal hierarchical spectrum (left), the smaller mass difference inferred from solar 
neutrino oscillations separates the lightest neutrino from the next-to-lightest, whereas the
larger mass difference inferred from atmospheric neutrino oscillations separates the latter from
the heaviest neutrino. In the inverted hierarchical spectrum (right) this is reversed.}
\label{fig:numass}
\end{figure}

The big bang relic neutrinos originate from the decoupling of the weak interactions
when the universe was about one second old and the primordial plasma had a temperature
of about one MeV, much larger than the possible neutrino masses. 
Neglecting late-time, small-scale gravitational clustering, to which we come
later, and in the absence of appreciable lepton asymmetries,\footnote{Large neutrino mixing 
inferred from oscillation experiments ensures the validity of the neglection of chemical 
potentials\cite{Lunardini:2000fy,Dolgov:2002ab,Wong:2002fa,Abazajian:2002qx}.} 
their phase space distribution is therefore 
predicted to be given by the homogenous and isotropic relativistic Fermi-Dirac distribution, 
$f_0(p)=1/(1+\exp (p/T_{\nu,0}))$, where $p$ is the modulus of the comoving three-momentum $\mathbf p$ and 
$T_{\nu,0}=(4/11)^{1/3}\,T_{\gamma,0}=1.95$~K is today's neutrino temperature.
Correspondingly, the gross properties of the C$\nu$B are tightly related to the 
properties of the well-measured CMB and are therefore to be considered as rather firm
predictions. Their present number density, 
\begin{eqnarray}
\label{nudens}
\underbrace{\bar n_{\nu_i\,0} 
                       =  
\bar n_{\bar\nu_i\,0}}_{{\rm C}\nu{\rm B}}
                       = \frac{3}{22}\, 
                       \underbrace{\bar n_{\gamma\,0}}_{\rm CMB}
                       = 56\ {\rm cm}^{-3}
\,,
\end{eqnarray}
when summed over all neutrino types $i=1,2,3$, is large and comparable to the one of the CMB, 
$\sum_i (\bar n_{\nu_i\,0}+ \bar n_{\bar\nu_i\,0}) = (9/11)\,\bar n_{\gamma\,0}$. 
Their present average three-momentum, on the other hand, is very small, 
\begin{eqnarray}
\underbrace{\bar p_{\nu_i\,0} = \bar p_{\bar\nu_i\,0}}_{ {\rm C}\nu{\rm B}} 
= 
3\, \bigl( \frac{4}{11}\bigr)^{1/3}\,\underbrace{T_{\gamma\,0}}_{\rm CMB} 
= 5\times 10^{-4}\ {\rm eV}
\,.
\end{eqnarray}
Correspondingly, at least two of the relic neutrino mass eigenstates are 
non-relativistic today ($m_{\nu_i}\gg \bar p_{\nu_i\,0}$), independently of whether 
neutrinos masses have a normal hierarchical or inverted hierarchical pattern 
(cf. Fig.~\ref{fig:numass}). 
These neutrinos are subject to gravitational clustering into gravitational 
potential wells due to existing CDM and baryonic structures, possibly causing
the local neutrino number density to be enhanced relative to the standard value~(\ref{nudens})
and the momentum distribution to deviate from the relativistic Fermi-Dirac distribution.  

\begin{figure}[t]
\vspace{-0.43cm} 
\begin{center}
         \mbox{\epsfig{file=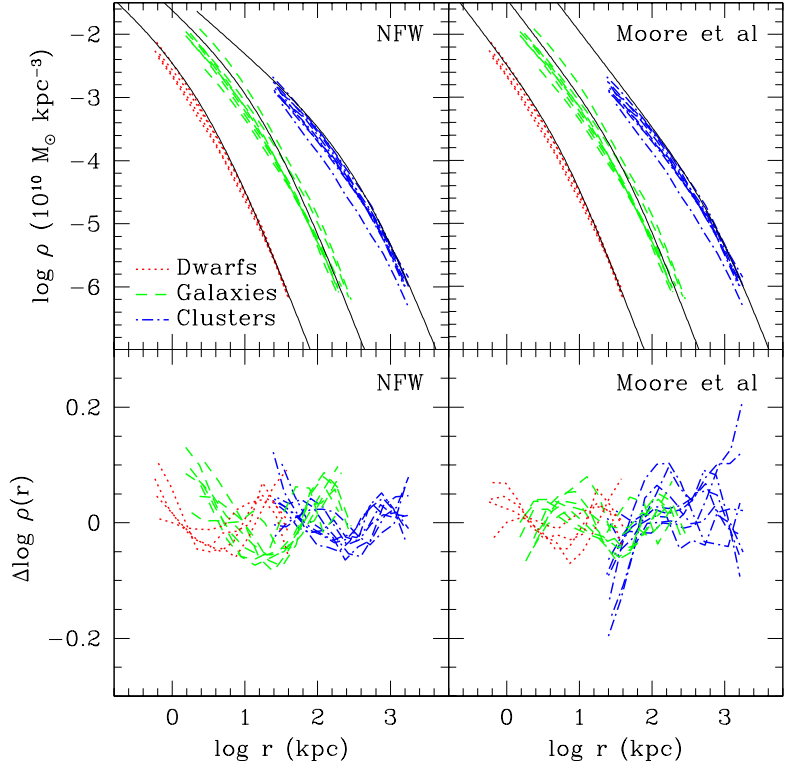,width=11cm,clip=}}
\end{center}
\caption[]{CDM mass density profiles inferred from $N$-body simulations of $\Lambda$CDM 
models\cite{Navarro:2003ew}.}
\label{fig:cdmprofiles}
\end{figure}

This question can be studied quantitatively as follows\cite{Singh:2002de,Ringwald:2004np}. 
First of all, in the context
of a flat $\Lambda$CDM model, the neutrino contribution to the total dark matter density is always
a small perturbation (cf. Table~\ref{tab:recipe}). Correspondingly, the CDM mass density $\rho_m$
dominates in the gravitational potential $\phi$. Secondly, the neutrinos will have negligible 
gravitational interactions with each other. Therefore, one can make use 
of the CDM halo profiles from high-quality, pure $\Lambda$CDM simulations 
(cf. Fig.~\ref{fig:cdmprofiles}) 
and study
the evolution of the neutrino phase space distribution $f_{\nu_i} (\mathbf{x},\mathbf{p},\tau)$
in the corresponding gravitational potential wells. This distribution depends 
on the comoving distance ${\mathbf x} = {\mathbf r}/a(t)$, its conjugate momentum 
${\mathbf p} = a m_{\nu_i}\, \dot{\mathbf x}$, and the conformal time $d\tau = dt /a(t)$,
$a$ being the cosmological scale factor.   
Technically speaking, one has then to solve the Vlasov, or collisionless Boltzmann equation,
\begin{eqnarray}
\label{eq:vlasov}
\frac{D f_{\nu_i}}{D \tau} \equiv
\frac{\partial f_{\nu_i}}{\partial \tau} + \dot{\mathbf x} 
\cdot \frac{\partial f_{\nu_i}}{\partial {\mathbf x}} \underbrace{-a m_{\nu_i} \nabla \phi}_{\dot{\mathbf p} }
\cdot \frac{\partial f_{\nu_i}}{\partial {\mathbf p}} = 0 
\,,
\end{eqnarray}
with the Poisson equation
\begin{eqnarray}
\nabla^2 \phi = 4 \pi G a^2  \underbrace{\bigl(
\rho_m({\mathbf x}, \tau) 
- \overline\rho_m(\tau)\bigr)}_{\delta_m(\mathbf{x},\tau)\,\overline\rho_m(\tau)}
\end{eqnarray}
relating the gravitational potential $\phi$ to the CDM density fluctuation $\delta_m$ with respect
to the physical mean $\bar{\rho}_m$. For a given CDM halo profile, e.g. the one advocated by 
Navarro, Frenk, and White (NFW)\cite{Navarro:1995iw,Navarro:1996he},    
\begin{eqnarray}
\rho_{m}(r) = \frac{\rho_s}{(r/r_s) (1 + r/r_s)^2}
\,, 
\end{eqnarray}
where the parameters $r_s$ and $\rho_s$ are basically determined by the halo's virial mass
$M_{\rm vir}$, the Vlasov equation~(\ref{eq:vlasov}) may be solved numerically by tracking 
the trajectories of neutrinos in phase space, starting from initial conditions corresponding to the  
homogeneous and isotropic Fermi-Dirac distribution 
($N$-one-body simulations)\cite{Ringwald:2004np}. The initial redshift can be taken as 
$z=3$, since, at higher redshifts, a sub-eV neutrino has too much thermal velocity to cluster 
efficiently.  

\begin{figure}[t]
\vspace{-.5cm}
\begin{center}
         \mbox{\epsfig{file=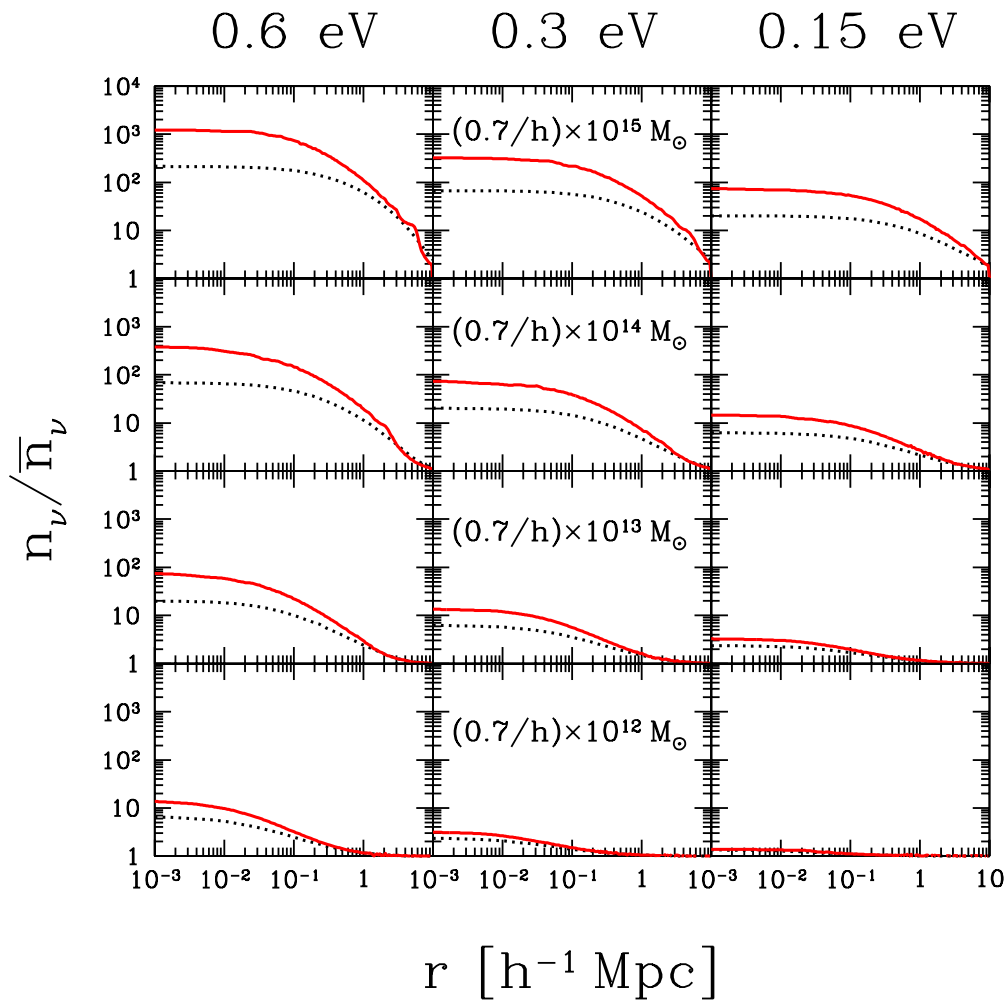,width=14cm,clip=}}
\end{center}
\caption[]{Neutrino number density profiles, normalized to their cosmological mean\cite{Ringwald:2004np}, 
calculated with $N$-one-body simulations (solid) and with the linear approximation (dotted), respectively.}
\label{fig:ovdenspanel}
\end{figure}

A comprensive and exhaustive comparative study\cite{Ringwald:2004np} has revealed the following 
results (cf. Fig.~\ref{fig:ovdenspanel}). First of all, a flattening
of the neutrino profiles at small radii is observed, in distinction to the CDM profiles 
(cf. Fig.~\ref{fig:cdmprofiles})).
This can be understood in terms of neutrino free-streaming. Secondly, the clustering is 
considerably improved towards larger CDM halo virial masses and/or larger neutrino masses. 
In the inner part ($\lwig 100$~kpc) of a massive galaxy cluster like e.g. the nearby ($\sim 15$~Mpc) 
Virgo cluster ($M_{\rm vir}\sim 10^{15}\,M_\odot$), the relic neutrino density $n_\nu$ can be larger than its
cosmological mean $\bar n_\nu$ by a factor of $\sim 10\div 1000$, if we take into account
the full range of possible neutrino masses for the heaviest neutrinos, 
$m_\nu = 0.04\div 0.6$~eV (cf. Fig.~\ref{fig:numass}). Much more moderate 
clustering, $n_\nu/\bar n_\nu\sim 1\div 20$, is obtained for 
ordinary galaxies like the Milky Way ($M_{\rm vir}\sim 10^{12}\,M_\odot$) 
in their central region ($\lwig 10$~kpc). 
We note in passing that a previously\cite{Singh:2002de} exploited semi-analytical linear method for solving
the Vlasov equation~(\ref{eq:vlasov}), which consists of replacing $\partial f/\partial \mathbf{p}$ 
by  $\partial f_0/\partial \mathbf{p}$, systematically underestimates the neutrino overdensities
over nearly the whole range of halo and neutrino masses considered here (cf. Fig.~\ref{fig:ovdenspanel}).     

\begin{figure}[t]
\vspace{-1.5cm}
\begin{center}
         \mbox{\epsfig{file=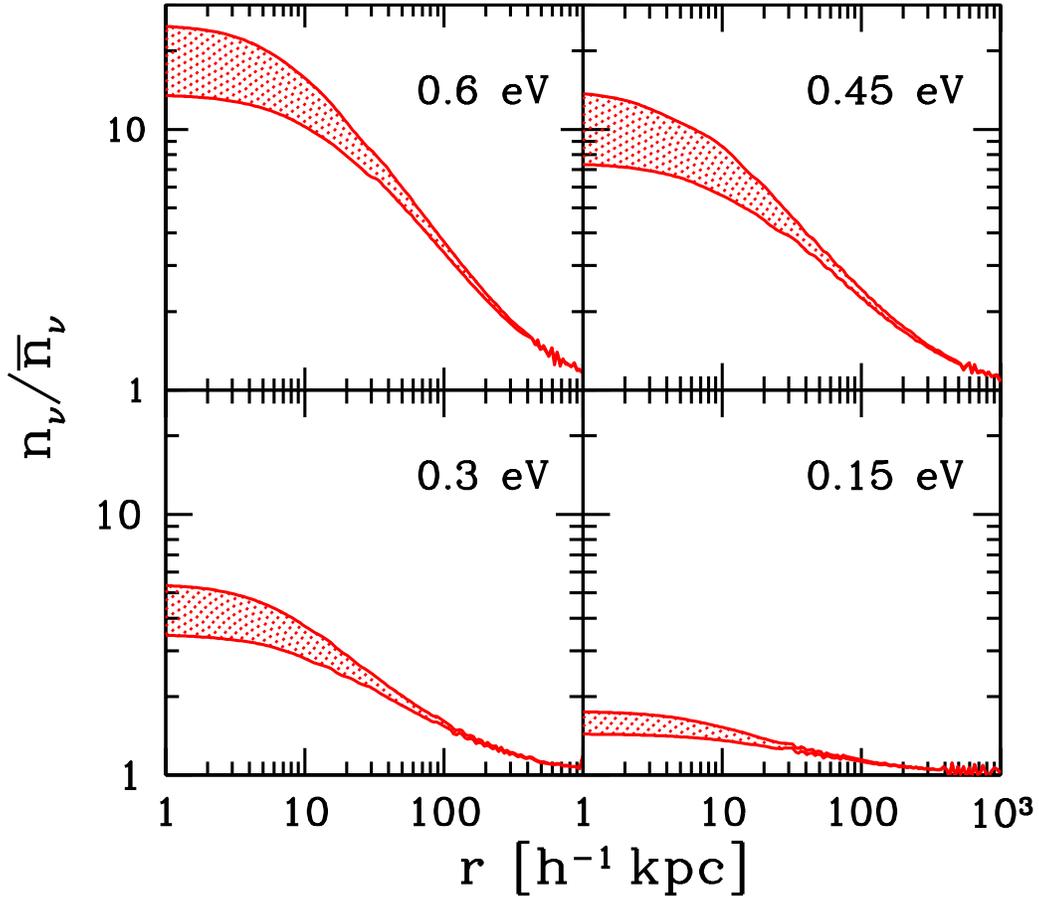,width=14cm,clip=}}
\end{center}
\caption[]{Neutrino density profiles for the Milky Way\cite{Ringwald:2004np}, obtained via 
$N$-one-body simulations from the MWnow (top curve in each plot) and the NFWhalo run (bottom 
curve in each plot).}
\label{fig:ovdensmw}
\end{figure}

For an accurate determination of the relic neutrino number density and momentum distribution
in the Earth's local neighbourhood in the Milky Way, at a distance $r_\oplus\sim 8$~kpc from the 
galactic center, one needs, in principle, to know the gravitational
potential over the history of the Milky Way, i.e. the complete assembly history. 
In the absence of that information, one may consider two extreme cases, with the true 
behavior somehow lying in-between: (i) working with the present day Milky Way mass 
distribution\cite{Dehnen:1996fa,Klypin:2001xu} (MWnow),
assuming it to be static (in physical coordinates), and (ii) exploiting the NFW halo (NFWhalo) that would
have been there, had baryon compression---which is thought to lead to the formation of the 
galactic bulge and disk---not taken place. The possible ranges of overdensities, $\sim 1\div 20$, 
in the Milky Way are illustrated 
in Fig.~\ref{fig:ovdensmw}. 

\begin{figure}[t]
\vspace{-1.5cm}
\begin{center}
         \mbox{\epsfig{file=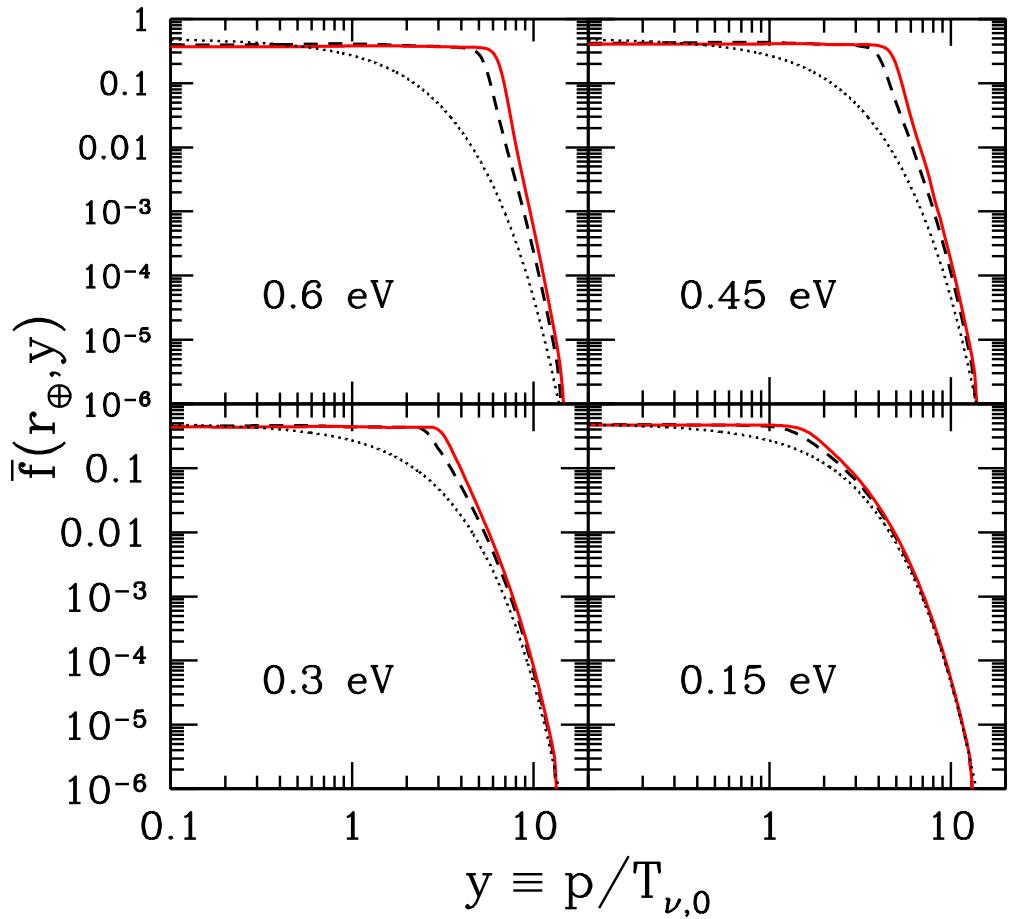,width=14cm,clip=}}
\end{center}
\caption[]{Momentum distribution of the relic neutrinos in the local neighbourhood of 
the Earth\cite{Ringwald:2004np}, 
obtained from the MWnow (solid) and NFWhalo (dashed) run, approaching for large momenta the
relativistic Fermi-Dirac distribution (dotted).}
\label{fig:momdist}
\end{figure} 

The final momentum distribution at $r_\oplus$ is found to be almost isotropic, with
mean radial velocity $\langle v_r\rangle\approx 0$ and second moments that satisfy 
approximately the relation $2\langle v_r^2\rangle \approx \langle v_T^2\rangle$. 
The coarse-grained phase space densities $\bar f(r_\oplus,p)$ in Fig.~\ref{fig:momdist} 
are flat at low momenta, with a common value of nearly 1/2, have a turning point at about the
escape momenta $p_{\rm esc}\equiv m_{\nu} v_{\rm esc} \equiv m_\nu \sqrt{2 |\phi(r_{\oplus})|}$, 
and quickly approach the Fermi-Dirac distribution for larger momenta.
Note, that the results displayed in Fig.~\ref{fig:momdist} not only satisfy, but, up to $p_{\rm esc}$, 
nearly completely saturate the general phase space bound\cite{Lynden-Bell:1966bi,Tremaine:1979we} 
$\bar f \leq {\rm max}(f_0)=1/2$.   
The corresponding semi-degenerate state can only be 
made denser by filling in states above $p_{\rm esc}$. 
In order to attain even higher densities, one must appeal to non-standard theories\cite{Dolgov:2005proc}.

\section{\label{howdetect}How to Detect?}

The gravitational infall of the relic neutrinos into CDM halos, discussed in the last section, 
might be significant 
for its influence on the non-linear (i.e. not so large-scale, wave number $k\gwig 1$~Mpc$^{-1}$) matter 
power spectrum\cite{Abazajian:2004zh,Hannestad:2005prep}, which will be determined in the upcoming 
weak gravitational lensing campaigns. In contrast to the evidences 
for the relic neutrinos from BBN, CMB, and the linear (i.e. large-scale, $k\lwig  1$~Mpc$^{-1}$) matter 
power spectrum 
mentioned in the Introduction, this gravitational inference will test the
presence of relic neutrinos at low redshift, i.e. near to the present epoch. 
In this section, we will concentrate on other detection techniques, which are also sensitive to the 
present C$\nu$B, which, however, are based instead on weak interaction scattering processes involving the
relic neutrinos either as a beam or as a target\cite{Ringwald:2004np}. 

\begin{table}[t]
\caption[]{Relic neutrino properties\cite{Ringwald:2004np} as relevant for flux detection.}
\label{tab:deBroglie}
  \small 
\begin{center}
\begin{tabular}{@{}l|l|l|l} 
\hline & $\frac{n_\nu}{\bar{n}_\nu}$  & $\lambdabar = 
\frac{1}{\langle p \rangle}$   
 & $\langle v\rangle$  
\\ 
\hline
MWnow     
\\ $m_{\nu}=$  
\\ \hline
$0.6 \ {\rm eV}$ &  20   & 
$2.3\times 10^{-2}$~cm   & 
 $1.4\times 10^{-3}$  \\ 
$0.45 \ {\rm eV}$  &  10   & 
$2.9\times 10^{-2}$~cm  & 
 $1.5\times 10^{-3}$  \\ 
$0.3 \ {\rm eV}$  &  4.4  & 
$3.7\times 10^{-2}$~cm  & 
 $1.8\times 10^{-3}$  \\ 
$0.15 \ {\rm eV}$ &  1.6   & 
$4.1\times 10^{-2}$~cm   & 
 $3.2\times 10^{-3}$  \\ 
\hline
NFWhalo    
\\ $m_{\nu}=$ 
\\ \hline
$0.6 \ {\rm eV}$ & 12  &  
$2.7\times 10^{-2}$~cm   & 
 $1.2\times 10^{-3}$  \\ 
$0.45 \ {\rm eV}$  &  6.4  & 
$3.4\times 10^{-2}$~cm  & 
 $1.3\times 10^{-3}$  \\ 
$0.3 \ {\rm eV}$ &  3.1   & 
$3.9\times 10^{-2}$~cm   & 
 $1.7\times 10^{-3}$  \\ 
$0.15 \ {\rm eV}$ &  1.4   &  
$5.9\times 10^{-2}$~cm   & 
$2.2\times 10^{-3}$  \\ 
\hline 
\end{tabular}
\end{center}
\end{table}

\subsection{\label{flux}Flux detection}

The Earth is moving through the almost isotropic (cf. last section) relic neutrino background. 
In this subsection, we consider coherent elastic scattering of the corresponding relic neutrino flux 
off target matter in a terrestrial detector. In this connection, it is an important observation that
the low average momentum of relic neutrinos corresponds to a de Broglie wavelength  
of macroscopic dimension, 
\begin{eqnarray} 
\lambdabar = 1/\langle p\rangle = 0.12\ {\rm ~cm}/\langle p/T_{\nu,0}\rangle
\,.
\end{eqnarray}
Correspondingly, one may envisage\cite{Shvartsman:sn,Smith:jj} scattering processes in which many target atoms 
of atomic mass $A$ 
act coherently over a macroscopic volume $\lambdabar^3$, yielding an enhancement of the  
elastic scattering rate by the huge factor  
\begin{eqnarray} 
\frac{N_A}{A}\,\rho_{\rm t}\,\lambdabar^3 
\simeq {6\times 10^{18}}\,\left( \frac{100}{A}\right)  
\left( \frac{\rho_{\rm t}}{{\rm g/cm^3}}\right) 
\left( \frac{\lambdabar}{0.1\ {\rm cm}}\right)^3   
\,,
\end{eqnarray}
where $N_A$ is Avogadro's number and $\rho_{\rm t}$ is the target mass density,  
compared to case where neutrinos are elastically scattered coherently only on  
the individual nuclei of the target.
\begin{figure}[t]
\begin{center}
         \mbox{\epsfig{file=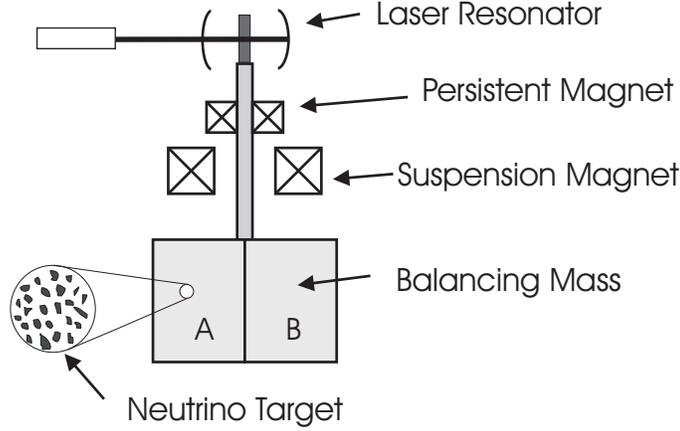,bbllx=80pt,bblly=337pt,bburx=586pt,bbury=662pt,%
width=9cm,clip=}}
\end{center}
\caption[]{Cavendish-type torsion balance for relic neutrino detection\cite{Hagmann:1998nz,Hagmann:1999kf}.}
\label{fig:torsionbalance}
\end{figure} 
A terrestrial target of linear size $r_{\rm t}<\lambdabar$  will 
therefore experience a neutrino wind induced acceleration\cite{Shvartsman:sn,Smith:jj,Duda:2001hd}
\begin{eqnarray} 
a_{\rm t} &\simeq & \sum_{\nu,\bar\nu}\  
\underbrace{n_{\nu}\,v_{\rm rel}}_{\rm flux}\ 
\frac{4\pi}{3}\, N_A^2\, \rho_{\rm t}\,  r_{\rm t}^3 
\  
\sigma_{\nu N}\,  
\underbrace{2\,m_\nu\,v_{\rm rel}}_{\rm mom.\, transfer} 
\nonumber \\[-.5ex]
\label{eq:accel}
&\simeq & 
{2\times 10^{-28}}\   
{\frac{\rm cm}{{\rm s}^{2}}}\ 
\left( \frac{n_\nu}{\bar n_\nu}\right)  
\left( \frac{10^{-3}\,c}{v_{\rm rel}}\right) 
\left( \frac{\rho_{\rm t}}{{\rm g/cm^3}}\right) 
\left( \frac{r_{\rm t}}{\lambdabar}\right)^3 
\,,
\end{eqnarray}  
where  $\sigma_{\nu N}\simeq G_F^2\, m_\nu^2/\pi$ is the elastic  
neutrino--nucleon cross section, and  
$v_{\rm rel}=\langle |\mathbf{v} - \mathbf{v}_\oplus|\rangle$ is the mean 
velocity of the relic neutrinos  
in the rest system of the  
detector. Here, $v_\oplus\simeq 7.7\times 10^{-4}\,c$ 
denotes  
the velocity of the Earth through the Milky Way. For Majorana neutrinos, 
the acceleration is further suppressed,  
in comparison with~(\ref{eq:accel}), by a factor of 
$(v_{\rm rel}/c)^{2}\simeq 10^{-6}$  
for an unpolarized and $v_{\rm rel}/c\simeq 10^{-3}$ for a polarized target, respectively.

What are the prospects to measure such small accelerations? 
Presently, conventional Cavendish-type torsion balances routinely reach 
$10^{-13} \ {\rm cm \ s}^{-2}$. Possible improvements with currently available 
technology to a sensitivity of $\gwig 10^{-23} \ {\rm cm \ s}^{-2}$ have been 
proposed\cite{Hagmann:1998nz,Hagmann:1999kf} (cf. Fig.~\ref{fig:torsionbalance})\footnote{
Such improvements would also be very interesting in the connection of the search 
for deviations from Newton's law at small distances and possible significant improvements on the bounds of 
the size of extra space-like dimensions\cite{Eidelman:2004wy}.}. 
However, even such an improved sensitivity is still off the prediction~(\ref{eq:accel}) by at 
least three orders of magnitude,  
as an inspection of the currently allowed range of local relic neutrino 
overdensities  in Table~\ref{tab:deBroglie} reveals.   
Therefore, an observation of this effect will not be possible within the upcoming decade. 
But it can still be envisaged in the not-so-distant future, say, within thirty to forty years. 
Note, in this context, that the acceleration~(\ref{eq:accel}) can be improved still by a considerable amount 
by using foam-like\cite{Shvartsman:sn} or laminated\cite{Smith:jj} materials.  In this way one may exploit 
a target size much larger than $\lambdabar$, while still avoiding destructive interference.  
Alternatively, grains of size $\sim\lambdabar$ could be randomly embedded 
(with spacing $\sim \lambdabar$) 
in a low density host material\cite{Smith:1991mm,Smith:sy}. Advances in nanotechnology may 
be very welcome in this connection. For the case of Majorana neutrinos, flux detection via 
mechanical forces will remain a real challenge, however.

\begin{table}[t]
\caption[]{Planned and projected accelerator beams and their interaction rates 
with the relic neutrinos\cite{Ringwald:2004np}.}\label{table:accelerators}
  \small 
\begin{center}
\begin{tabular}{@{}l|c|r|r|r||r} 
\hline accel. & $N$    &$E_N$   & $L$  & $I$ & 
$\frac{R_{\nu A}}{\left[\frac{n_\nu}
{\bar n_\nu}\,\frac{m_\nu}{\rm eV}\right]}$ \\  
 &   & [TeV]  &  [km] &  [A] & [yr$^{-1}$] \\ 
\hline
 & $p$  & $7$  &  $26.7$ &  $0.6$   & 
$2\times 10^{-8}$\\
LHC            &            &                           &       
&         &              \\
 & Pb  & $574$  & $26.7$ &   $0.006$      &    
$1\times 10^{-5}$\\
\hline
 & $p$ &  $87.5$  & $233$ & $0.06$  & 
$2\times 10^{-7}$\\ 
VLHC &   &    &  &   & \\ 
 & Pb  &  $7280$  & $233$ & $0.0006$  & 
$1\times 10^{-4}$\\ 
\hline
ULHC & $p$  &  $10^7$  & $40\,000$ & $0.1$  
& $10$ \\ 
\hline
\end{tabular} 
\end{center}
\end{table}

\subsection{\label{targetdetection}Target detection}

The weak interaction cross sections are rapidly growing with energy, at least at 
center-of-mass energies below the $W$- and $Z$-resonances. 
In this subsection, we study the question whether the scattering of extremely energetic particles 
(accelerator beams or cosmic rays) off the relic neutrinos as a target has promising prospects
for C$\nu$B detection. 

\begin{figure}[t]
\begin{center}
         \mbox{\epsfig{file=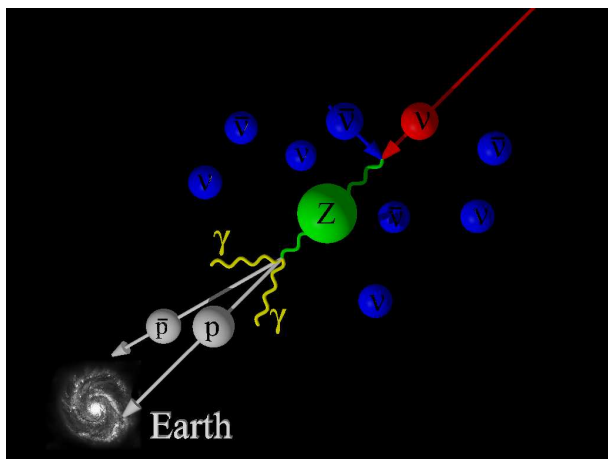,width=8cm,clip=}}
\end{center}
\caption[]{Resonant annihilation of an extremely energetic cosmic neutrino  off a relic anti-neutrino
into a $Z$-boson\cite{Fodor:2002cc}.}
\label{fig:zburst}
\end{figure}

\subsubsection{Exploiting accelerator beams}

For a beam of particles ${}^A_Z N$, with energy $E_N$, charge $Z\,e$, length $L$,  and current $I$,  
the interaction rate with the relic neutrinos is of 
order\cite{Muller:1987qm,Melissinos:1999ew,Weiler:2001wc} 
\begin{eqnarray}
\label{eq:acc_rate}
       { R_{\nu\, {}^A_Z N}} &\simeq & 
\sum_{\nu,\bar\nu}\ n_\nu\,\sigma_{\nu\, {}^A_Z N}\,L\ I/(Z\,e)
\\[-.5ex]  \nonumber &\simeq & 
       {{ 2\times 10^{-8}}\ { {\rm yr}^{-1}}}
\ 
\left( \frac{n_\nu}{\bar n_\nu}\right)
       \left( \frac{m_{\nu}}{{\rm eV}}\right)
 \frac{A^2}{Z}\,
       \left( \frac{E_N}{{\rm 10\ TeV}}\right)
 \left( \frac{L}{100\ {\rm km}}\right)
       \left( \frac{I}{0.1\ {\rm A}}\right)
\,.
\end{eqnarray}
In view of the currently allowed range of local relic neutrino overdensities, $\sim 1\div 20$,  
and the beam parameters of the next and next-to-next generation of accelerators---the Large Hadron Collider 
(LHC) 
and the Very Large Hadron Collider (VLHC), respectively---the expected rate~(\ref{eq:acc_rate}) 
is clearly too small to give rise to an observable effect in the foreseeable 
future (cf. Table~\ref{table:accelerators}). 
Even with an Ultimate Large Hadron Collider (ULHC) designed to accelerate protons to energies above 
$10^7 \ {\rm TeV}$ 
in a ring of ultimate circumference  
$L\simeq 4\times 10^4 \ {\rm km}$ around the Earth,\footnote{Note that such an accelerator, 
in the  collider mode,  will 
probe the ``intermediate'' scale $(M_{\rm EW}\,M_{\rm GUT})^{1/2}\sim 10^{10} \ {\rm GeV}$ between
the electroweak scale $M_{\rm EW}\sim 1 \ {\rm TeV}$ 
and the scale of grand unification $M_{\rm GUT}\sim 10^{17} \ {\rm GeV}$. 
The intermediate scale is exploited in many schemes of 
supersymmetry breaking and in seesaw mechanisms for neutrino masses.}  it seems very difficult to establish
the interactions with the relic neutrinos, although they occur at a rate of more than one event per year 
(cf. Table~\ref{table:accelerators}): 
elastic scattering of the beam particles with the relic neutrinos---one of the contributions to the 
rate~(\ref{eq:acc_rate})---will be next to impossible to detect 
because of the small momentum transfers 
involved ($\sim 1 \ {\rm GeV}$ at $E_N\sim 10^7 \ {\rm TeV}$).  
A very promising alternative is to consider a heavy ion beam, and 
to exploit the contribution of the inverse beta decay reaction,
\begin{equation}
\label{eq:inversebeta}
{}^A_Z N + \nu_e \to {}^A_{Z+1} N + e^- 
\,,
\end{equation}
to the rate~(\ref{eq:acc_rate}).
This reaction  
changes the charge of the nucleus, causing it to follow 
an extraordinary trajectory and finally to exit the machine
such that it becomes susceptible
to detection\cite{Melissinos:1999ew,Zavattini:unpubl}. 
A detection of this reaction would also clearly
demonstrate that a neutrino was involved in the scattering.

\subsubsection{Exploiting cosmic rays}

\begin{figure}[t]
\begin{center}
         \mbox{\epsfig{file=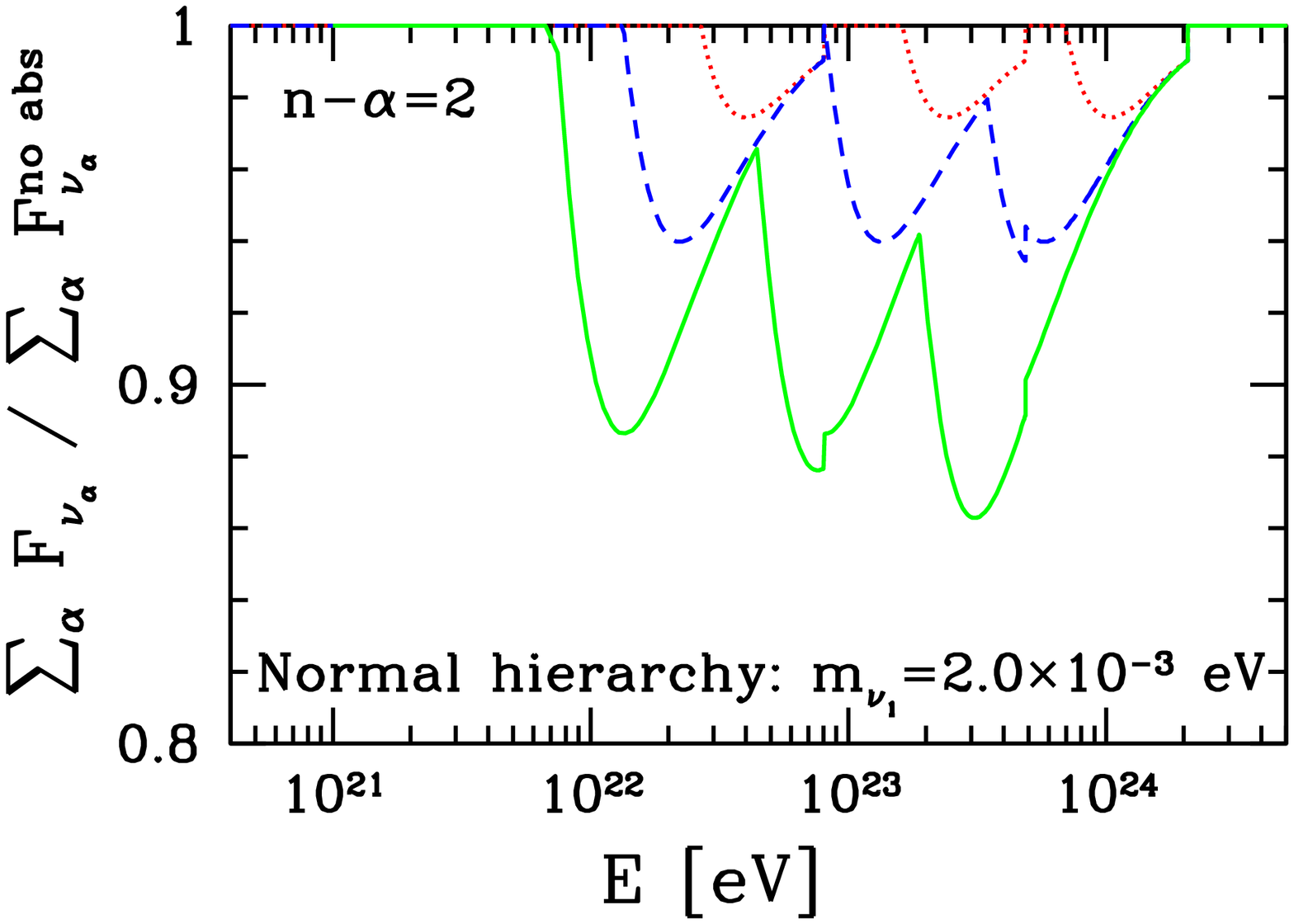,bbllx=20pt,bblly=225pt,%
bburx=570pt,bbury=608pt,width=7.5cm,clip=}}
\hfill
         \mbox{\epsfig{file=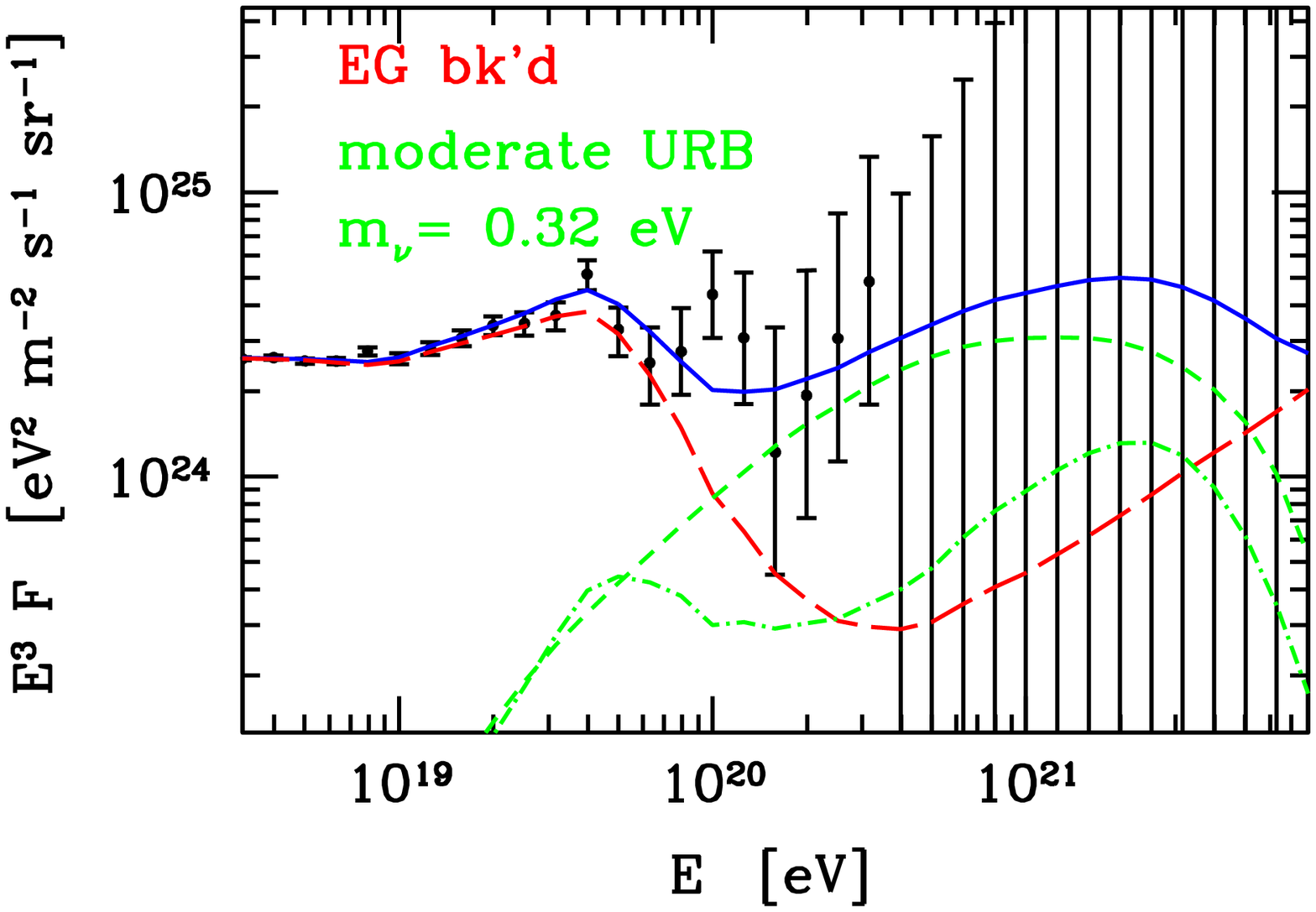,bbllx=20pt,bblly=225pt,%
bburx=570pt,bbury=608pt,width=7.5cm,clip=}}
\end{center}
\caption[]{Signatures of resonant annihilation of extremely energetic cosmic neutrinos off relic neutrinos:
absorption dips (left) in the EEC$\nu$ flux\cite{Eberle:2004ua}, for an isotropic source distribution  
out to redshift $z_{\rm max}=2,5,10$ (from upper to lower curves), 
and emission features ($Z$-bursts) (right) 
in the EEC ray (EECR) flux\cite{Fodor:2002hy} flux in the form of photons (short-dashed) 
and protons (dashed-dotted), eventually overcoming the ordinary extragalactic EECR flux (long dashed), 
which suffers from the GZK cutoff.}
\label{fig:signatures}
\end{figure}

In the foreseeable future, before the commissioning of the ULHC, target detection of the 
relic neutrinos has to rely on extremely energetic cosmic rays. Indeed, 
the resonant annihilation of extremely energetic cosmic neutrinos (EEC$\nu$s) off 
relic anti-neutrinos into $Z$-bosons (cf. Fig.~\ref{fig:zburst}), occuring at the resonance energies
\begin{equation}
E^{\rm res}_{\nu_i}=\frac{m_Z^2}{2m_{\nu_i} }\simeq 4\times 10^{21}\,
\left( \frac{\rm eV}{m_{\nu_i} }\right)
 \ {\rm eV},
\end{equation} 
offers unique opportunities for relic neutrino detection.   
On the one hand, one may search for absorption 
dips\cite{Weiler:1982qy,Weiler:1983xx,Roulet:1993pz,Yoshida:1997ie,Eberle:2004ua,Barenboim:2004di,Barenboim:2005proc} 
in the EEC$\nu$ spectrum at the resonant energies (cf. Fig.~\ref{fig:signatures} (left)), 
on the other hand, one may look for 
emission 
featu\-res\cite{Fargion:1999ft,Weiler:1999sh,Yoshida:1998it,Fodor:2001qy,Fodor:2002hy,Gelmini:2004zb} 
($Z$-bursts) 
as protons or photons with energies spanning a decade or more above 
the predicted Greisen--Zatsepin--Kuzmin (GZK) cutoff\cite{Greisen:1966jv,Zatsepin:1966jv} at 
$E_{\rm GZK}\simeq 4\times 10^{19}$~eV 
(cf. Fig.~\ref{fig:signatures} (right)). 
This is the energy beyond which the CMB is absorbing to nucleons due to 
resonant photopion production. Indeed, the association of $Z$-bursts with 
the mysterious post-GZK cosmic rays observed\cite{Takeda:1998ps} by the Akeno Giant Air Shower Array (AGASA) 
is a controversial\cite{Kalashev:2001sh,Gorbunov:2002nb,Semikoz:2003wv} 
possibility\cite{Fargion:1999ft,Weiler:1999sh,Yoshida:1998it,Fodor:2001qy,Fodor:2002hy,Gelmini:2004zb} 
(cf. Fig.~\ref{fig:signatures} (right)). 

\begin{figure}[t]
\begin{center}
         \mbox{\epsfig{file=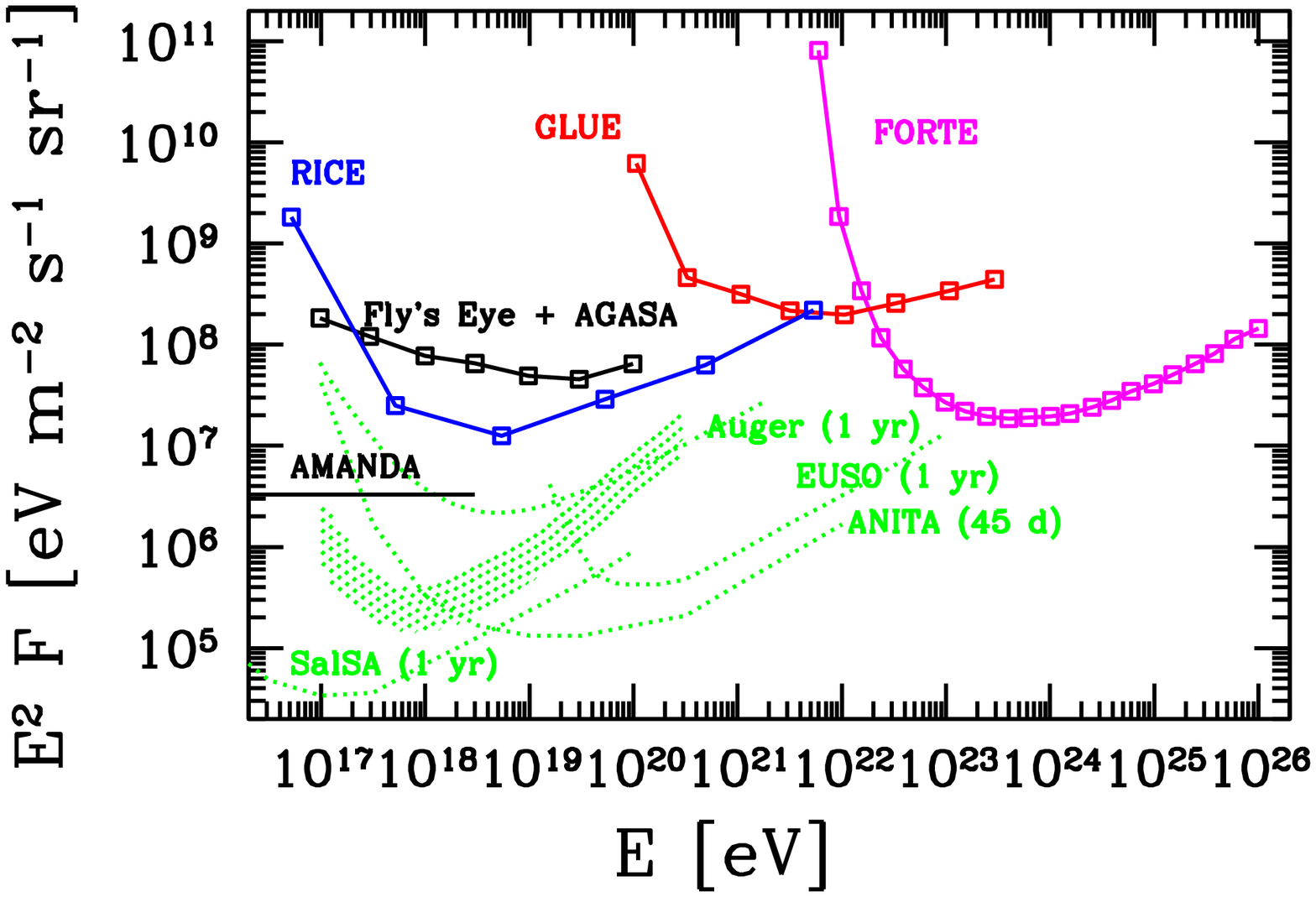,bbllx=20pt,bblly=225pt,%
bburx=570pt,bbury=608pt,width=7.5cm,clip=}}
\hfill
         \mbox{\epsfig{file=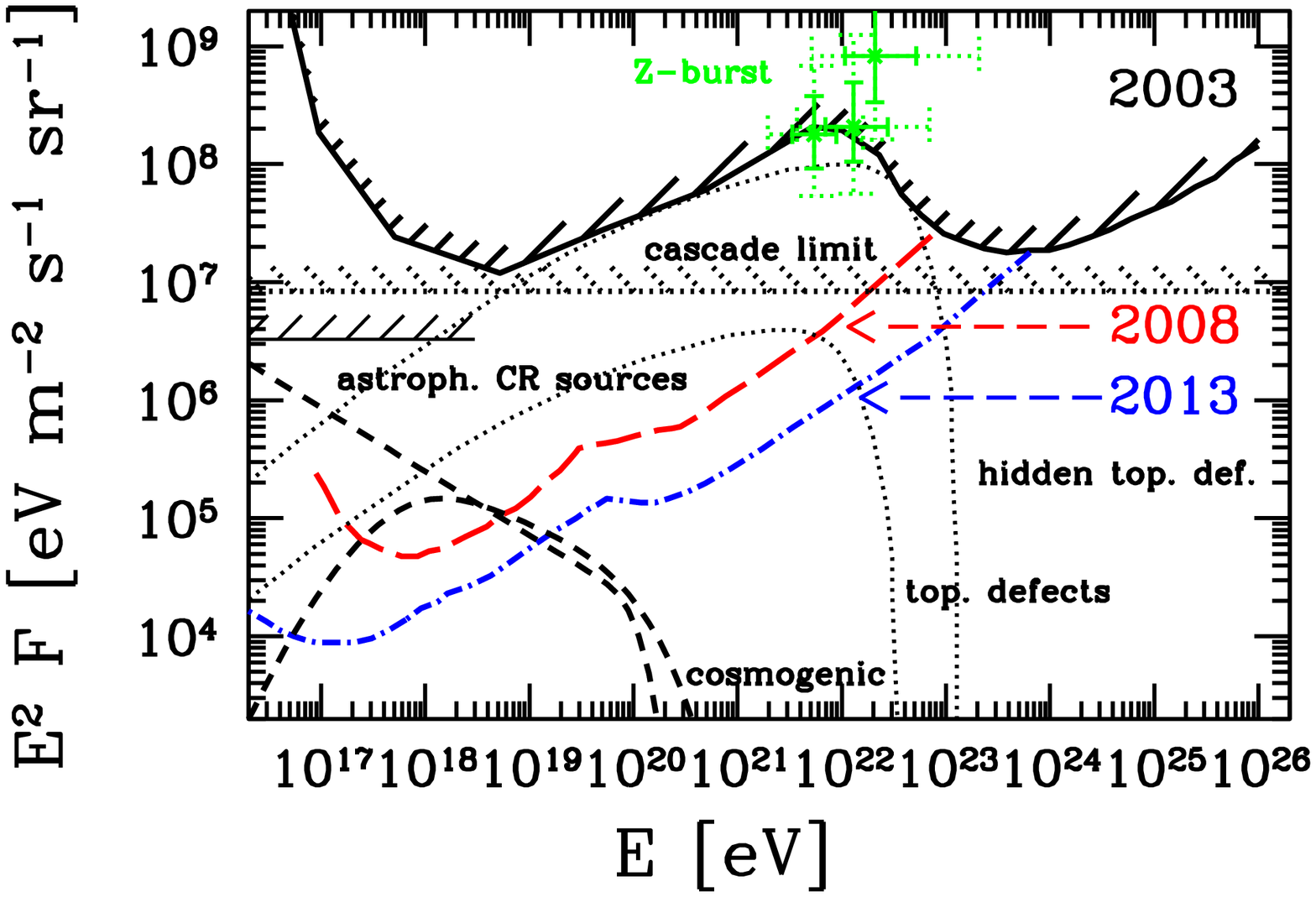,bbllx=20pt,bblly=225pt,%
bburx=570pt,bbury=608pt,width=7.5cm,clip=}}
\end{center}
\caption[]{EEC$\nu$ 
fluxes\cite{Eberle:2004ua,Ringwald:2005prep,Berezinsky:2005proc,%
Halzen:2005proc,Learned:2005proc,McKay:2005proc,Weiler:2005proc}: 
current limits and projected sensitivities of dedicated 
experiments (left), as well as prospects within the next decade (2008 and 2013) 
and theoretical predictions (right).}
\label{fig:prospects}
\end{figure}

\begin{figure}[t]
\begin{center}
         \mbox{\epsfig{file=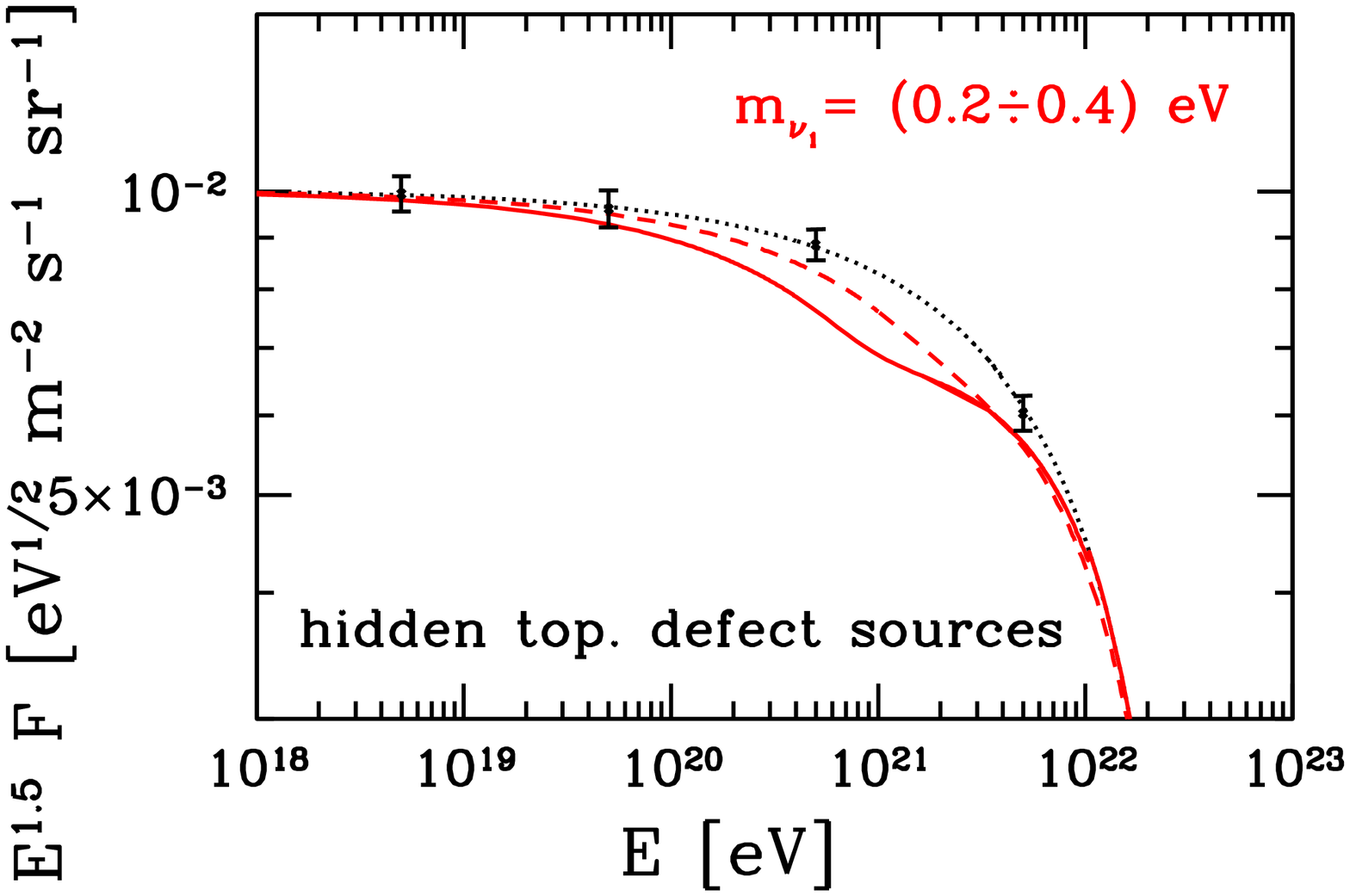,bbllx=20pt,bblly=225pt,%
bburx=570pt,bbury=608pt,width=7.5cm,clip=}}
\hfill
         \mbox{\epsfig{file=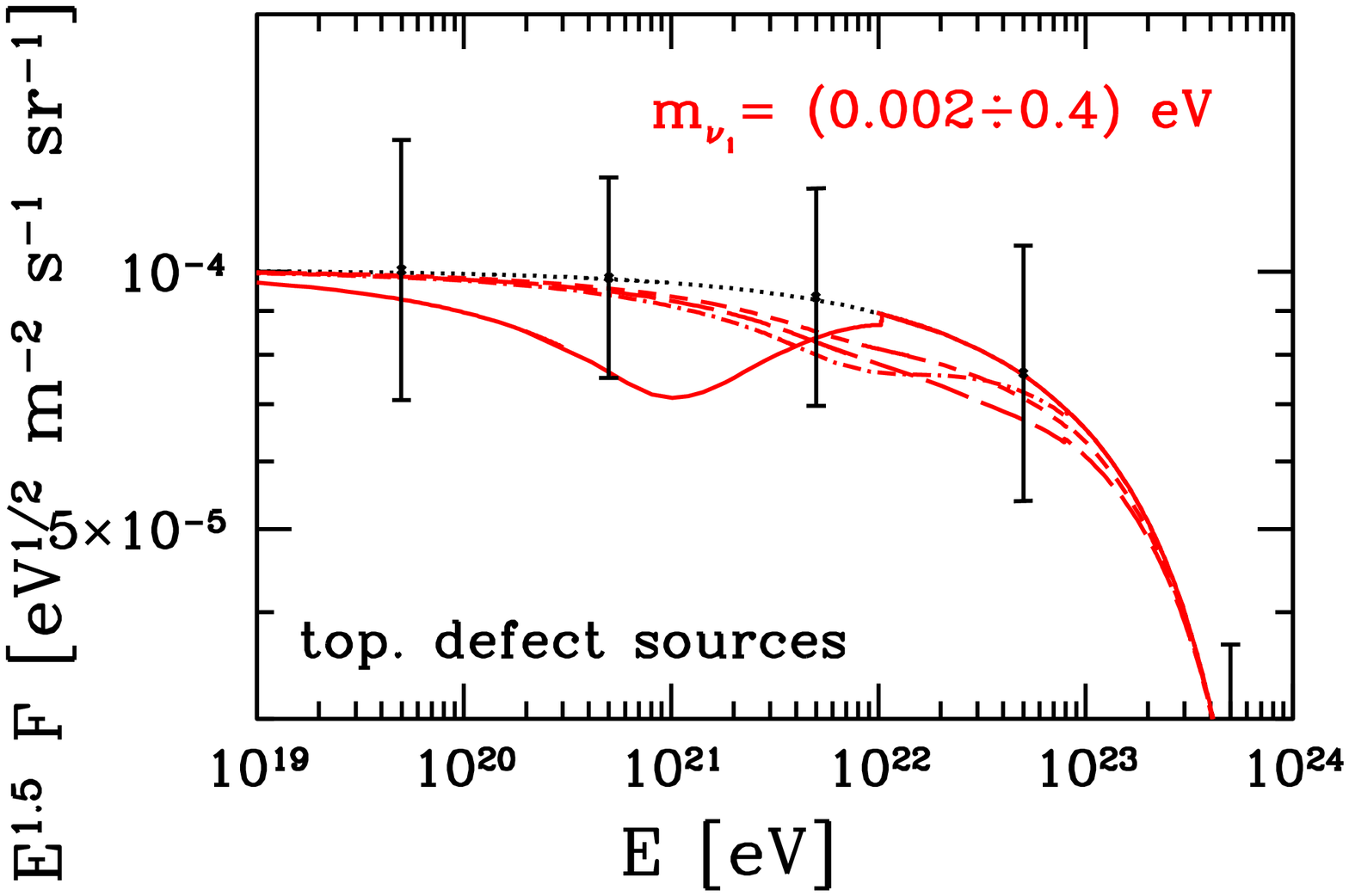,bbllx=20pt,bblly=225pt,%
bburx=570pt,bbury=608pt,width=7.5cm,clip=}}
\end{center}
\caption[]{Relic neutrino absorption dips in the EEC$\nu$ flux\cite{Eberle:2004ua} 
from hidden topological defect sources (left) and topological defect sources (right),  
with projected error bars in the year 2013.}
\label{fig:prospectsrnabs}
\end{figure}

Presently planned neutrino detectors (Pierre Auger Observatory\cite{Auger}, 
IceCube\cite{IceCube}, ANITA\cite{ANITA}, EUSO\cite{EUSO}, 
OWL\cite{OWL}, and SalSA\cite{Gorham:2001wr})
operating in the energy regime above $10^{21} \ {\rm eV}$  appear to 
be sensitive enough to lead us, within the next decade (cf. Fig.~\ref{fig:prospects}), 
into an era of relic neutrino absorption 
spectroscopy\cite{Eberle:2004ua,Barenboim:2004di,Barenboim:2005proc}  (cf. Fig.~\ref{fig:prospectsrnabs}), 
provided that the EEC$\nu$ flux at the resonant energies is close to current observational
bounds and the neutrino mass is sufficiently large, $m_\nu\gwig\, 0.1 \ {\rm eV}$. 
In this context it is important to note, that absorption spectroscopy is predominantly sensitive 
to the relic neutrino background at early times, with the depths of the absorption dips determined 
largely by the higher number densities at large redshifts ($z\gg 1$) (cf. Fig.~\ref{fig:signatures} (left)). 
Since neutrinos do not cluster significantly until after $z\lwig 2$, clustering at recent times 
can only show up as secondary dips with such minimal widths in energy\cite{Reiter:unpubl}  
that they do not seem likely to be resolved by planned observatories. 

\begin{figure}[t]
\vspace{-.5cm}
\begin{center}
         \mbox{\epsfig{file=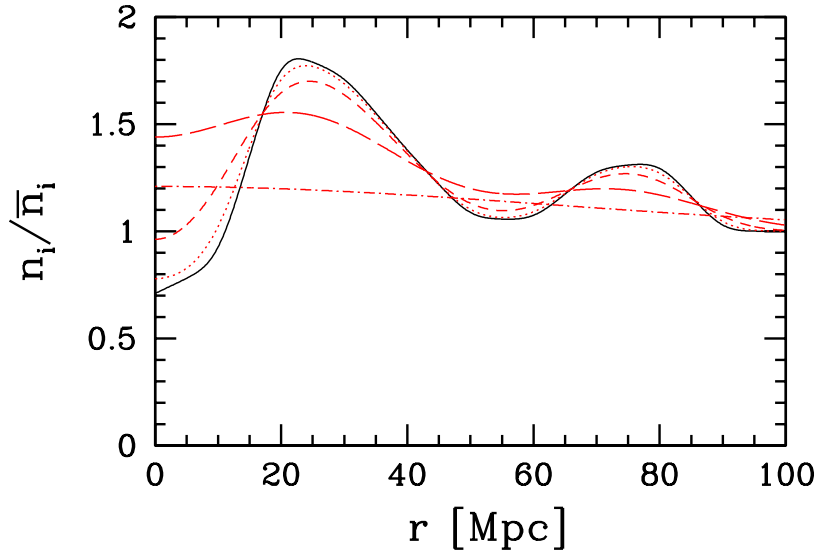,width=11.cm,clip=}}
\end{center}
\caption[]{``Large scale'' overdensities ($i=\nu,{\rm CDM}$) in the 
local universe\cite{Ringwald:2004np}, with the Milky Way at the origin.  
The black (solid) line corresponds to the local CDM distribution\cite{Fodor:2002hy}, 
inferred from peculiar velocity measurements\cite{daCosta:1996nt},  
smeared over the surface of a sphere with radius $r$.  
The dotted line is the neutrino overdensity for
$m_{\nu} = 0.6 \ {\rm eV}$, short dash $0.3 \ {\rm  eV}$, long dash 
$0.15 \ {\rm eV}$, and dot-dash $0.04 \ {\rm eV}$.}
\label{fig:ovdenslocaluni}
\end{figure}

On the other hand, emission spectroscopy is directly sensitive to the 
relic neutrino content 
of the local universe ($z\lwig 0.01 \Leftrightarrow 
r_{\rm GZK} \lwig 50 \ {\rm Mpc}$).
However, since the neutrino density contrasts approximately track those of 
the underlying CDM above the neutrino 
free-streaming scale, it is clear that there cannot be a substantial neutrino overdensity over the 
whole GZK volume ($\sim r_{\rm GZK}^3$).  Indeed, the estimated
neutrino overdensity in our local GZK zone, with a $\sim 5 \ {\rm Mpc}$ smoothing,  is always  $\lwig 2$ 
(cf. Fig.~\ref{fig:ovdenslocaluni}). 
Hence, the overall emission
rate cannot be significantly enhanced by gravitational clustering.  

\begin{figure}[t]
\vspace{-1.5cm}
\begin{center}
         \mbox{\epsfig{file=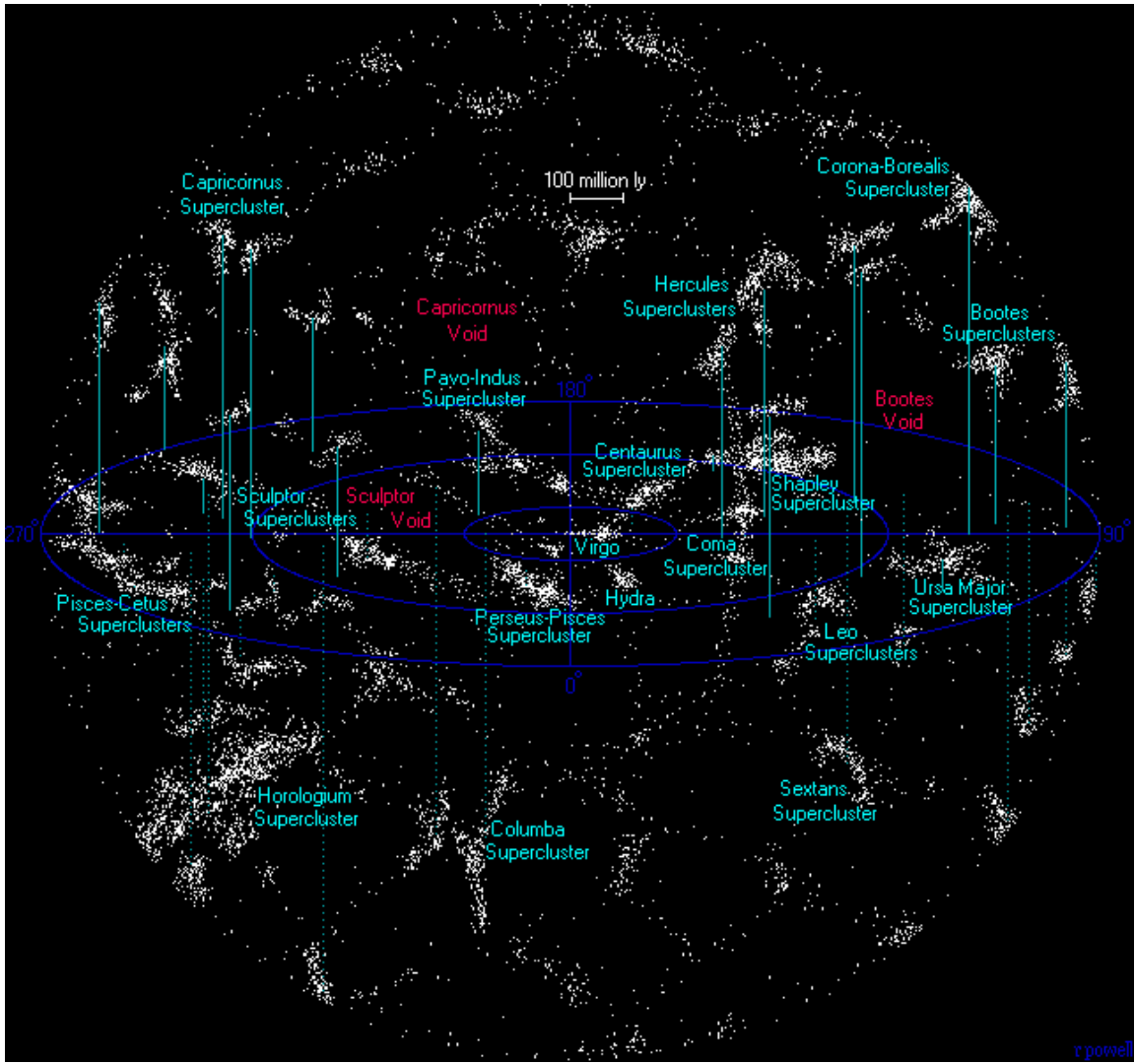,width=15.25cm,clip=}}
\end{center}
\caption[]{Various massive galaxy superclusters\cite{bib:supercluster} 
in our ``vicinity''. The Milky Way is at the center of the coordinate system.}
\label{superclusters}
\end{figure}

Nevertheless, it seems worthwhile to contemplate about ``relic neutrino tomography'' of the 
local universe\cite{Ringwald:2004np,Ringwald:2005prep}.  
Specifically, one may exploit the fact that there are several
galaxy clusters ($\! \gwig  10^{14} M_{\odot}$), such as
Virgo (distance $\sim 15 \ {\rm Mpc}$) and Centaurus ($\sim 45 \ {\rm Mpc}$),
within the GZK zone (cf. Figure~\ref{superclusters}) within which we expect significant
neutrino clustering (cf. Figure~\ref{fig:ovdenspanel}). 
One could then search for directional dependences in the emission events as a signature of 
EEC$\nu$ annihilating on relic anti-neutrinos (and vice versa)\cite{Ringwald:2004np}. 
For example, the angular resolution of AGASA, $\sim 2^{\circ}$, is already sufficient 
to resolve the internal structures of, say, the Virgo cluster
($M_{\rm vir} \sim  10^{15} M_{\odot}$)
which spans some $10^{\circ}$ across the sky.
Using the $N$-one-body clustering results in Figure~\ref{fig:ovdenspanel},
the average neutrino overdensity along the line of 
sight towards and up to Virgo is estimated to be 
$\sim 45$ and $\sim 5$ for $m_{\nu} = 0.6 \ {\rm eV}$ and
$0.15 \ {\rm eV}$ respectively, given an angular resolution of 
$\sim 2^{\circ}$.  The corresponding increases in the number of events
coming from the direction of the Virgo cluster relative to the unclustered
case, assuming an isotropic distribution of EEC$\nu$ sources,
 are given roughly by the same numbers, since protons originating
from $\sim 15 \ {\rm Mpc}$ away arrive at Earth approximately unattenuated.
The numbers improve to $\sim 55$ and $\sim 8$ respectively with a finer
$\sim 1^{\circ}$ angular resolution. In the most optimistic case of a EEC$\nu$ flux
near to the current observational bound (cf. Fig.~\ref{fig:prospects} (right)) and a neutrinos mass 
$\gwig 0.1$~eV, EUSO will not only find evidence for the absorption dips, but also 
for the enhanced emission from the direction of Virgo due to $Z$-bursts\cite{Ringwald:2005prep}.

\section{\label{conclusions}Conclusions}

At present, BBN, CMB, and the large-scale matter power spectrum provide the
only observational evidence for the big bang relic neutrinos, at least in the early stages
of the cosmological evolution. 
A more direct, weak interaction based detection of the C$\nu$B near 
the present epoch may proceed in the following chronological order by measuring 
\begin{itemize}
\item[i)] absorption dips in EEC$\nu$ spectra and $Z$-bursts in EECR spectra; 
\item[ii)] macroscopic forces through coherent elastic scattering of relic neutrinos 
          off target material in Cavendish-type torsion balances;
\item[iii)] interactions of extremely energetic particles from terrestrial accelerator
          beams with the relic neutrinos as a target. 
\end{itemize}
Unfortunately, an immediate and guaranteed direct detection does not appear to be feasible. 
Although the search for signatures of EEC$\nu$ annihilation off the relic neutrinos can start 
right now, its success entirely rests on the existence of an EEC$\nu$ flux at the resonance energies. 
The sensitivity of Cavendish-type torsion balances has still to be improved by at least thirteen orders
of magnitude for a detectable signal, which postpones this detection techniques probably beyond 
the year of retirement of the author, $\gwig 2025$. Finally, an appreciable rate of beam particles with the 
relic neutrinos requires an  Ultimate Large Hadron Collider around the Earth with a beam energy 
$E_{\rm beam}\gwig 10^7$~TeV, which almost certainly will never be built. In the meantime, we can hope
to see the late-time relic neutrinos through their gravitational effects in the 
not-so-large-scale, non-linear part of the matter power spectrum measured by weak gravitational lensing.

\section{\label{acknowl}Acknowledgments}

I would like to thank Birgit Eberle, Zoltan Fodor, Sandor Katz, Tom Weiler, and 
Yvonne Wong for the nice collaboration on different aspects of relic neutrino 
detection.

\end{document}